\documentstyle[11pt,iau211,twoside,epsf]{article}
\begin{document}
\title{Parallaxes of Brown Dwarfs at USNO}
\author{Hugh C. Harris, Conard C. Dahn, Frederick J. Vrba, Harry H. Guetter,
Blaise Canzian, Arne A. Henden, Stephen E. Levine, Christian B. Luginbuhl,
Alice K. B. Monet, David G. Monet, Jeffery R. Pier, Ronald C. Stone,
Richard L. Walker}
\affil{U. S. Naval Observatory, P.O. Box 1149, Flagstaff, AZ 86002-1149}

\begin{abstract}
Trigonometric parallaxes have been measured by Dahn et al.\ (2002)
for 28 cool dwarfs and brown dwarfs, including 17 L dwarfs and three
T dwarfs.  Broadband CCD and near-IR photometry ($VRIz^*JHK$) have been
obtained for these objects and for 24 additional late-type dwarfs.
These data have been supplemented with astrometry and photometry from
the literature, including parallaxes for the brighter companions of
ten L and two T dwarfs.  The absolute magnitudes and colors are
reviewed here.  The $I-J$ color and the spectral type are both good
predictors of absolute magnitude for late-M and L dwarfs.
$M_J$ becomes monotonically fainter with $I-J$ color and with spectral
type through late-L dwarfs, then brightens for early-T dwarfs.
In contrast, the $J-K$ color correlates poorly with absolute magnitude
for L dwarfs.  Using several other parameters from the literature
(Li detection, H$\alpha$ emission strength, projected rotation velocity,
and tangential velocity), we fail to uncover any measurable parameter
that correlates with the anomalous $J-K$ color.
\end{abstract}

\section{Introduction}
Parallaxes of late-M, L, and T dwarfs have been measured from images
taken over the past several years with the 1.55 m Strand astrometric
telescope at the U.S. Naval Observatory.  The present results have been
published recently by Dahn et al. (2002).  The parallaxes are necessary
for many purposes, including comparison with evolutionary models,
analyzing the kinematics, determining ages, determining temperatures,
and identifying outlying objects (perhaps binaries or young objects).
When faint companions to bright stars with Hipparcos parallaxes are
included, there are now nearly 30 field L dwarfs with measured and
published distances, but only five T dwarfs published to date.
In this paper, we discuss two applications of the parallax data
to our understanding of L dwarfs.

\section{Predicting Absolute Magnitudes}
In Figure 1, the absolute magnitude $M_J$ is plotted.
($M_J$ is used because it scales more closely with luminosity
than $M_I$ or $M_K$ do.)  The five known binary pairs of L dwarfs
are labelled in the middle panel -- they are plotted after correcting
their magnitudes from the combined light to the individual components.
The figure shows that both spectral type and $I-J$ color can be used
to predict $M_J$ for M7 through L8 spectral types.  
\begin{figure}[t]
\plotone{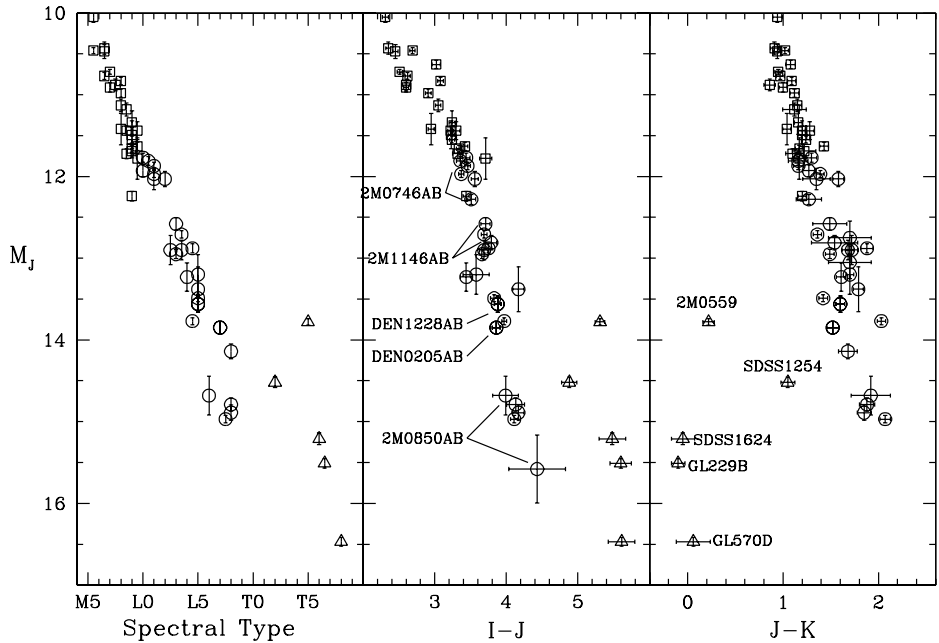}
\caption{The absolute magnitude $M_J$ plotted against spectral type
and color.  M dwarfs are shown by squares, L dwarfs by circles,
and T dwarfs by triangles.
Binary L dwarfs are labelled in the middle panel,
and T dwarfs are labelled in the righthand panel.}
\end{figure}
The scatter in the $J-K$ color of L dwarfs makes that color less
accurate for use in predicting $M_J$.
We need more data before drawing conclusions about T dwarfs.

\section{The Near-IR Colors of L Dwarfs}
The $JHK$ colors of L dwarfs are known to have a poor correlation
with spectral type (e.g. Leggett et al. 2002) and with absolute
magnitude and temperature (Kirkpatrick et al. 2000; Dahn et al. 2002).
The right-hand panel of Figure~1 reinforces that conclusion.
The $JHK$ colors are believed to be controlled by the amount of dust
in the atmospheres of these cool objects:  the distribution of dust
particle sizes, the vertical distribution of dust clouds,
and perhaps the existance of holes or bands between dust clouds
are likely to be relevant factors
(e.g. Ackerman \& Marley 2001; Burrows et al. 2002; Tsuji 2002).
\begin{figure}[hp]
\plotone{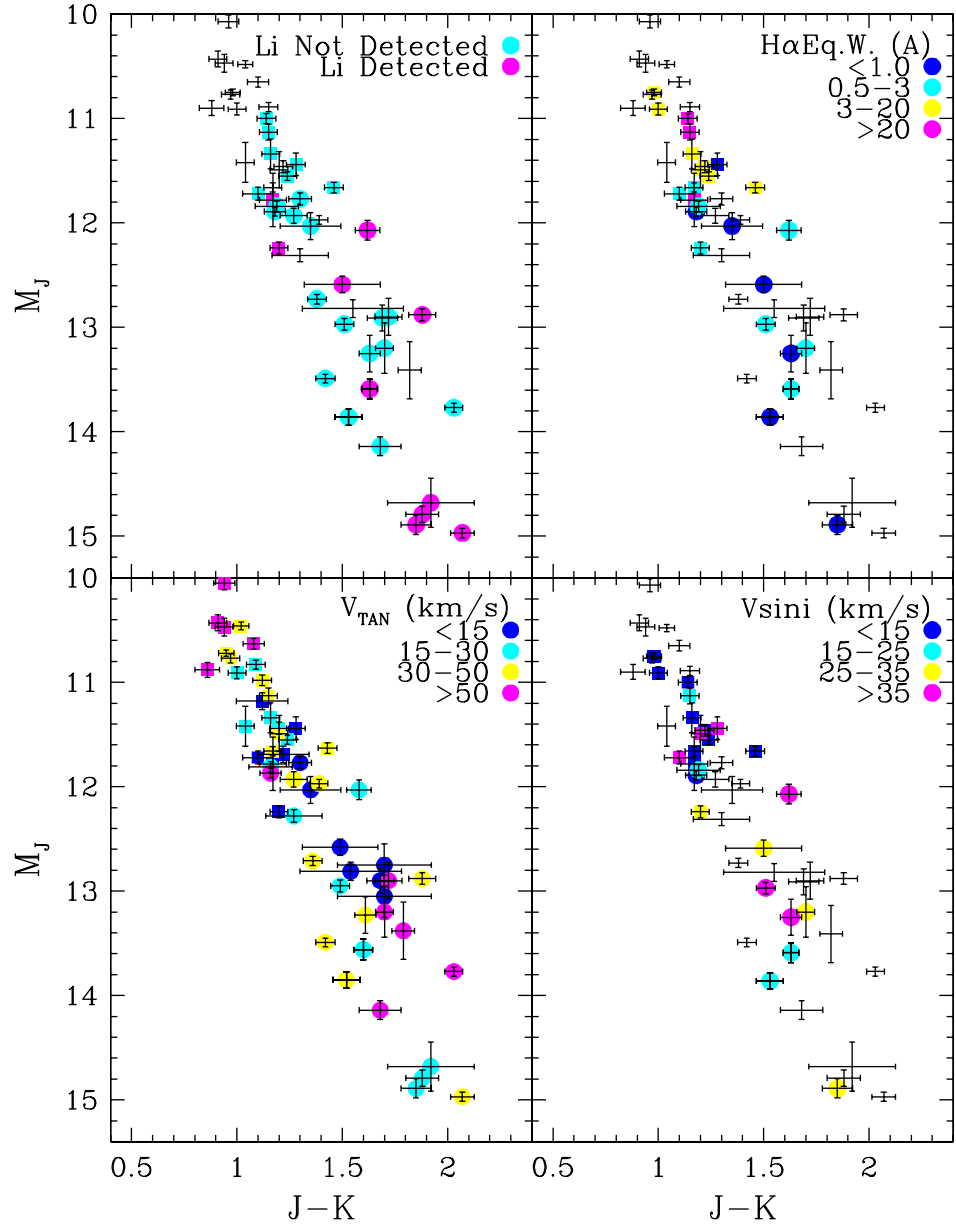}
\caption{The near-IR color-absolute magnitude diagram for late-M and
L dwarfs.  At a given absolute magnitude, the $J-K$ color shows little
correlation with any of these four parameters.}
\end{figure}

In an effort to understand what factor(s) cause dust to vary between
otherwise-similar objects, in Figure 2 we repeat the right-hand panel
of Figure 1 to show four potential correlations.
Figure 2 uses data from the literature
(Basri et al. 2000; Basri 2001; 
Kirkpatrick et al. 1999; 2000; 2001;
Mart\'{\i}n et al. 1999; Reid et al. 2002; Tinney \& Reid 1998)
for lithium detection, H$\alpha$ emission, tangential velocity,
and rotation velocity.  Some trends are already known:
lithium is detected more frequently at fainter absolute magnitudes,
while H$\alpha$ equivalent widths drop and rotation velocities increase.
However, based on the available data, there is no obvious tendency for
the $J-K$ color to correlate with any of these parameters at a given
absolute magnitude.  Perhaps several factors influence the dust content
of the atmospheres of L dwarfs and mask any one-parameter correlation.

\section{Future work}
We need data (parallax and other) for more L dwarfs.
We really need data for more T dwarfs.
Parallaxes for more T dwarfs will be coming from Tinney (this volume)
and from parallaxes with the USNO IR camera (Vrba et al.
in preparation).

\acknowledgments 
Many objects discussed here were first identified from the 2MASS and
SDSS surveys, and we are grateful to our colleagues for notifying
us of objects in advance of publication.

\end{document}